\documentclass[11pt]{article}

\usepackage{lineno}
\usepackage{tabularx}

\modulolinenumbers[1]

\usepackage{pstricks}

\usepackage{amsmath}
\usepackage{amssymb}
\usepackage{color}
\usepackage{soul}
\usepackage{graphicx}

\usepackage{setspace}

\begin{document}

\begin{center}
{\LARGE {\bf The Frictional Brachistochrone}}
\\[7mm]
\end{center}
{Shiva P. Pudasaini\\[3mm]
Technical University of Munich, School of Engineering and Design\\
Civil and Environmental Engineering,}
 Arcisstrasse 21, D-80333, Munich, Germany\\[2mm]
Kathmandu Institute of Complex Flows\\ Kageshwori Manohara - 3,
Bhadrabas, Kathmandu, Nepal\\[2mm]
{E-mail: shiva.pudasaini@tum.de}\\[7mm]
{\bf Abstract:} 
Here, I construct an elegant frictional brachistochrone for a mass point motion of a granular material with the Coulomb frictional energy dissipation that inherently includes the evolving path curvature. The simple model reveals several striking mechanical phenomena. It is applicable to any frictional particle. With increasing friction, the particle path becomes less and less curved until a straight brachistochrone is attained in the limit of sufficiently high friction. The existence of the straight-brachistochrone is phenomenal. Some potential industrial applications of the frictional brachistochrone are considered.

\section{Introduction}

The classical brachistochrone (Bernoulli, 1696) is one of the brilliant inventions in mathematics and physics. It applies to an ideal situation without energy dissipation. This provides a perfect example of perpetual cycloidal motion of a particle. The real world, however, is governed by the gravitational acceleration and the frictional energy dissipation along a curved path with evolving curvature. 
\\[3mm]
Several models exist for the brachistochrone considering friction (Ashby et al., 1975; Lipp, 1997; Hayen, 2005; Covic and Veskovic, 2008; Weisstein, 2022; Barsuk and Paladi, 2023). Some of them are simple, but do not include curvature of the path, while others, that include curvature, are complex. Based on the energy balance for a particle (or a mass point) moving along a curved path, I derive a simple, yet elegant equation for the fastest path, known as brachistochrone. This equation incorporates the Coulomb frictional energy dissipation and instantaneous curvature of the path. 
\\[3mm]
The newly developed frictional brachistochrone clearly demonstrates the significant to dominant effect of the Coulomb friction as it inherently includes the evolving path curvature. As the friction increases, brachistochrone tends to curve less and less. Consequently, in the limit, for sufficiently high friction, the brachistochrone tends to become a straight brachistochrone. 
The results are explained with respect to the mechanisms of the Coulomb frictional forces along the curved path of fastest travel time. Prospective applications of the frictional brachistochrone for the optimal design of the ski jump, toboggan run and the granular bulk transportation in channels in process engineering plants are discussed.

\section{The model}

The modelling process involves several steps including the energy balance along a curved path, constructing an expression for the particle velocity, minimization of the travel time, identification of the Lagrangian, and application of the Euler-Lagrange equation. This results in the first-ever, simple frictional brachistochrone for a mass point motion of a granular material. 

\subsection{Energy balance}

Consider the energy balance (per unit mass) along the particle path defined by its angle $\zeta = \zeta(s)$ in terms of the arc length $s$ including the change in the kinetic energy, gravitational potential energy, the Coulomb frictional energy dissipation due to the load normal to the slope and its enhancement due to the slope curvature, respectively (Pudasaini and Hutter, 2007):  
\begin{eqnarray}
\begin{array}{lll}
\displaystyle{d\left(\frac{1}{2}u^2\right) = g \sin\zeta ds - \mu g \cos\zeta ds - \mu\kappa u^2 ds}.
\label{Energy0}
\end{array}    
\end{eqnarray}
In (\ref{Energy0}), $u$ is the particle velocity along the slope, 
$g$ is the gravitational acceleration, 
 $\mu = \tan\delta$ is the Coulomb friction parameter with the basal friction angle $\delta$,
$\kappa = -\partial \zeta /\partial s = -\zeta'$ is the slope curvature, 
 and $ds$ is an element of the particle path. 
The energy balance (\ref{Energy0}) can be rewritten in the form of a simple differential equation:
\begin{eqnarray}
\begin{array}{lll}
\displaystyle{
\frac{d}{ds}\left(\frac{1}{2}u^2\right) = g \sin\zeta - \mu g \cos\zeta - \mu\kappa u^2 
}.
\label{Energy1}
\end{array}    
\end{eqnarray}
Since  $\displaystyle{\frac{ds}{dt} = u}$ and $\displaystyle{\frac{du}{dt} = u\frac{du}{ds} = \frac{d}{ds}\left(\frac{1}{2}u^2\right)}$,
(\ref{Energy1}) describes the mass point motion of the granular material along the curved path (Pudasaini and Hutter, 2007; Pudasaini and Krautblatter, 2022).
\\[3mm]
It is a simple fact from basic mathematics and physics, that the motion of a frictional particle down a curved path can be described appropriately by considering the path-fitted (curvilinear) coordinate system (Pudasaini and Hutter, 2007). Both the gravitational potential energy along the slope, and the Coulomb frictional energy dissipation that depends on the load (normal to the sliding path) as well as its enhancement due to the centrifugal force (normal to the sliding path) cannot be properly described by following the Cartesian coordinate system as they deviate substantially away from their natural states. 

\subsection{The particle velocity}

With the substitution $v = u^2$, and function definitions 
\begin{eqnarray}
\begin{array}{lll}
f(s) = 2\mu\kappa,\,\,\, g(s) = 2g(\sin\zeta - \mu\cos\zeta), 
\label{Energy1aa}
\end{array}    
\end{eqnarray}
(\ref{Energy1}) can be recast as:
\begin{eqnarray}
\begin{array}{lll}
\displaystyle{\frac{dv}{ds} + f(s)v-g(s) =0}.
\label{Energy1a}
\end{array}    
\end{eqnarray}
This is an ordinary differential equation that can be solved analytically for $v$, yielding:
\begin{eqnarray}
\begin{array}{lll}
\displaystyle{v = \exp\left(-\int_0^s f(\xi) d\xi \right)\int_0^s\exp\left(\int_0^\eta f(\chi) d\chi \right)g(\eta) d\eta}.
\label{Energy1b}
\end{array}    
\end{eqnarray}
Since $\kappa = -\partial\zeta/\partial s$, integrating the terms associated with $f$, and restoring its definition from (\ref{Energy1aa}), (\ref{Energy1b}) turns into
\begin{eqnarray}
\begin{array}{lll}
\displaystyle{v = 2g\exp\left[2\mu(\zeta(s)-\zeta(0)) \right]\int_0^s \exp\left[-2\mu(\zeta(\eta)-\zeta(0)) \right]
\left [\sin\zeta(\eta) - \mu\cos\zeta(\eta)\right] d\eta }.
\label{Energy1c}
\end{array}    
\end{eqnarray}
 With this, an expression for the particle velocity $u$ is obtained: 
\begin{eqnarray}
\begin{array}{lll}
\displaystyle{u = \sqrt{2g\exp\left[2\mu(\zeta(s)-\zeta(0)) \right]\int_0^s \exp\left[-2\mu(\zeta(\eta)-\zeta(0)) \right]
\left [\sin\zeta(\eta) - \mu\cos\zeta(\eta)\right] d\eta}}\,,
\label{Energy2d}
\end{array}    
\end{eqnarray}
which is a function of the arc length $s$ and the slope angle $\zeta$, and involves the Coulomb friction parameter $\mu$ together with the path curvature.

\subsection{The time minimization}

The time the particle takes while moving from position one $(s_1)$ to the next position $(s_2)$ along the path is given by
\begin{eqnarray}
\begin{array}{lll}
\displaystyle{t_{sp} 
= \int_{t_1}^{t_2} dt 
= \int_{s_1}^{s_2} \frac{dt}{ds} ds 
= \int_{s_1}^{s_2} \frac{1}{ds/dt} ds
= \int_{s_1}^{s_2} \frac{1}{u} ds
},
\label{Time1}
\end{array}    
\end{eqnarray}
where the path positions $s_1$ and $s_2$ correspond to the time $t_1$ and $t_2$, respectively.
With (\ref{Energy2d}), (\ref{Time1}) takes the form: 
\begin{eqnarray}
\begin{array}{lll}
\displaystyle{t_{sp} = \int_{0}^{s} \displaystyle{\frac{1}{\sqrt{2g\exp\left[2\mu(\zeta(\eta)-\zeta(0)) \right]
\int_0^\eta \exp\left[-2\mu(\zeta(\chi)-\zeta(0)) \right]
\left [\sin\zeta(\chi) - \mu\cos\zeta(\chi)\right] d\chi}}}\,{d\eta}
}.
\label{Time2}
\end{array}    
\end{eqnarray}
The main problem now is to minimize $t_{sp}$ by finding the stationary state of (\ref{Time2}) providing the path of the fastest travel time from position $s_1$ to position $s_2$, which is the definition of the brachistochrone (Bernoulli, 1696).
This time minimization is the most crucial aspect here that is achieved by the variational problem (Gelfand and Fomin, 1963; Hayen, 2005; Barsuk and Paladi, 2023) designed with a Lagrangian and the Euler-Lagrange formulation that I deal with below.

\subsection{The Lagrangian}

The integrand of (\ref{Time2}), as defined with $\mathcal L$:
\begin{eqnarray}
\begin{array}{lll}
\displaystyle{\mathcal L = {\frac{1}{\sqrt{2g\exp\left[2\mu(\zeta(s)-\zeta(0)) \right]\int_0^s \exp\left[-2\mu(\zeta(\eta)-\zeta(0)) \right]
\left [\sin\zeta(\eta) - \mu\cos\zeta(\eta)\right] d\eta}}}
},
\label{Lagrangian}
\end{array}    
\end{eqnarray}
is the Lagrangian of the system, which is a function of $s$ and $\zeta$. The Lagrangian functional $\mathcal L$ plays the pivotal role in constructing the brachistochrone. 

\subsection{The Euler-Lagrange equation}

The stationary point of (\ref{Time2}) is given by the Euler-Lagrange equation:
\begin{eqnarray}
\begin{array}{lll}
\displaystyle{ \frac{\partial \mathcal L}{\partial \zeta} - \frac{d}{ds}\left (\frac{\partial \mathcal L}{\partial \zeta'} \right) = 0.
}
\label{EulerLagrangian}
\end{array}    
\end{eqnarray}
Since the Lagrangian $\mathcal L$ is independent of $\zeta'$, (\ref{EulerLagrangian}) reduces to
\begin{eqnarray}
\begin{array}{lll}
\displaystyle{ \frac{\partial \mathcal L}{\partial \zeta} = 0.
}
\label{EulerLagrangian_1}
\end{array}    
\end{eqnarray}
So, there exists a constant $P$ (with the dimension of energy per unit mass) such that, after integration, (\ref{EulerLagrangian_1}) becomes:
\begin{eqnarray}
\begin{array}{lll}
\displaystyle{ \mathcal L = \frac{1}{\sqrt{P}}.
}
\label{EulerLagrangian_2}
\end{array}    
\end{eqnarray}
Applying (\ref{Lagrangian}) into (\ref{EulerLagrangian_2}) and simplifying, yields:
\begin{eqnarray}
\begin{array}{lll}
\displaystyle{\frac{P}{2g}\exp\left[-2\mu(\zeta(s)-\zeta(0))\right] = \int_0^s \exp\left[-2\mu(\zeta(\eta)-\zeta(0)) \right]
\left [\sin\zeta(\eta) - \mu\cos\zeta(\eta)\right] d\eta}.
\label{EulerLagrangian_3}
\end{array}    
\end{eqnarray}
This forms the basis for the frictional brachistochrone.

\subsection{The frictional brachistochrone}

Differentiating both sides of (\ref{EulerLagrangian_3}), after simplification, 
I obtain the frictional brachistochrone:
\begin{eqnarray}
\begin{array}{lll}
\displaystyle{
\frac{d\zeta}{ds} = -\frac{g}{P}\frac{1}{\mu}\left[ \sin\zeta(s) - \mu\cos\zeta(s)\right]
}.
\label{brachistochrone}
\end{array}    
\end{eqnarray}
The first order ordinary differential equation (\ref{brachistochrone}) explains the frictional brachistochrone in terms of the slope angle $\zeta$ as a function of the arc length $s$ as the particle moves along the (oriented) curved path. This is probably the most simple and elegant model equation for the frictional brachistochrone, architecting the fastest (minimum time) path of the frictional particle, while considering both the Coulomb frictional energy dissipation and the energy dissipation due to the curvature of the path. 
It is crucial to recognize that considering the system in the path-fitted coordinate with the inclusion of the curvature enabled the development of the graceful representation (\ref{brachistochrone}) for the frictional brachistochrone.
\\[3mm]
Now, for the ease of visualization, I transfer (\ref{brachistochrone}) to the Cartesian coordinate system by utilizing the rule 
\begin{eqnarray}
\begin{array}{lll}
\displaystyle{
dx = \cos\zeta ds, \,\,\,\,  dy = \sin\zeta ds, \,\,\,\, \frac{dy}{dx} = \tan\zeta}.
\label{brachistochrone-transform}
\end{array}    
\end{eqnarray}
After some elementary trigonometric operations and integration, with (\ref{brachistochrone-transform}), the frictional brachistochrone (\ref{brachistochrone}) takes the simple and pleasing form
\begin{eqnarray}
\begin{array}{lll}
\displaystyle{
\frac{dy}{dx} = \tan\left[\frac{g}{P}\frac{1}{\mu}\left(\mu x - y \right) \right]
}.
\label{brachistochrone-transform-1}
\end{array}    
\end{eqnarray}
Structurally, this is a phenomenal development, genuinely advancing the science of frictional brachistochrone.
\\[3mm]
In order to implement boundary conditions at the given end positions $s_1$ and $s_2$, I convert (\ref{brachistochrone-transform-1}) into a second order ordinary differential equation describing the frictional brachistochrone. Applying the arctangent on both sides, after differentiation with respect to $x$ and simplification, (\ref{brachistochrone-transform-1}) leads to:
\begin{eqnarray}
\begin{array}{lll}
\displaystyle{
y_{xx} = \lambda\left(1+y_x^2 \right)\left (\mu - y_x\right),
}
\label{brachistochrone-transform-11}
\end{array}    
\end{eqnarray}
where, $\displaystyle{y_{x} = \frac{dy}{dx} = \tan\left[\lambda\left (\mu x -y \right)\right],\,y_{xx} = \frac{d^2y}{dx^2}}$ and $\displaystyle{\lambda = \frac{g}{P}\frac{1}{\mu}}$, respectively. 
I call $\displaystyle{\widehat{g}_s = \frac{g}{P}}$ the scaled gravity.
It shows that, if the curvature is neglected, (\ref{brachistochrone-transform-11}) turns into $\displaystyle{\frac{dy}{dx} = \mu}$, which is a straight line, but not a brachistochrone. As I discuss at Section 3.2, this only applies to the limiting situation when the friction is sufficiently high. 
So, it is worth noting that for a frictional particle moving along a curved path, both the energy dissipations due to the classical Coulomb frictional force and the force induced by the curvature of the path must be considered for the brachistochrone to be physically meaningful. 
Neglection of the path curvature leads to an unphysical result.   
\\[3mm]
An exact, analytical solution to the ordinary differential equation (\ref{brachistochrone-transform-1}) is constructed, which reads:
\begin{eqnarray}
\begin{array}{lll}
\displaystyle{
\left(1+\mu^2\right)\left(K + \lambda x\right) = \mu \left(\lambda\left(\mu x -y\right)\right)-\ln\left [\mu\cos\left(\lambda\left(\mu x -y\right)\right)-\sin\left(\lambda\left(\mu x -y\right)\right)  \right],
}
\label{brachistochrone-transform-3}
\end{array}    
\end{eqnarray}
where, $K$ is a constant of integration that can be fixed with the boundary condition.

\section{The dynamics of the frictional brachistochrone}

\subsection{The friction control over the brachistochrone}

\begin{figure}[t!]
\begin{center}
  \includegraphics[width=14cm]{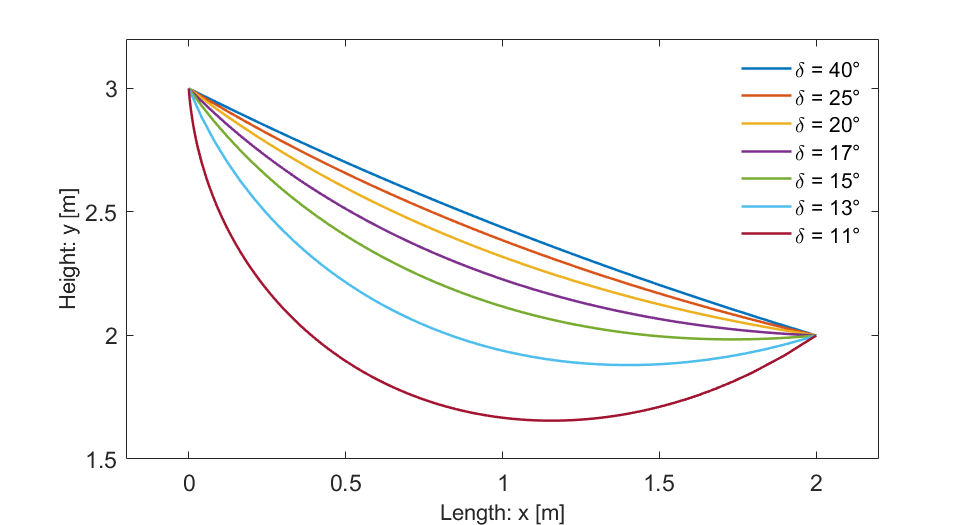}
  \end{center}
  \vspace{-5mm}
  \caption[]{The frictional brachistochrone with different energy dissipations associated with the friction
   angle $\delta$.}
   \label{Fig_1}
     \begin{picture}(3,3)
   \put(397.,109.7){\bf $\bullet$}
   \put(138.9,211.2){\bf $\bullet$}
    \put(397.,118.7){\bf $s_2$}
   \put(138.9,220.2){\bf $s_1$}
\end{picture}
\end{figure}

To disclose its dynamical behaviour, the solution to the frictional brachistochrone (\ref{brachistochrone-transform-11}) is presented in {Fig.\,\ref{Fig_1}} for $P =100$.
It reveals several interesting mechanical phenomena.
First, the reference analysis is performed for $\mu = 0.1944\,(\delta = 11^\circ)$, an extraordinarily low friction angle, akin to exceptionally smooth particles.
As the particle motion is triggered, first it descends along a path of higher slope where the particle accelerates swiftly. 
Then, as the particle continues to move sufficiently further the path descends slowly, where the particle decelerates. 
Yet, if the particle still possesses enough momentum, then, because of the slow decrease of the slope angle of the path, the particle can move along the path, for quite a while. 
\\[3mm]
Next, the analysis is performed with the increasing Coulomb friction parameter $\mu$. As the value of $\mu$ increases from its lower value ($\mu = 0.1944$) to higher values 
($\mu = 0.2309,\,0.2679,\,0.3057,\,0.3640,\,0.4663,\,0.8391$, corresponding to 
$\delta = 13^\circ, 15^\circ, 17^\circ, 20^\circ, 25^\circ, 40^\circ$), that represent highly frictional particles, the steepness of the path decreases, and it becomes straighter, now much earlier than with the lower friction parameter. This can be explained mechanically. This is so, because the energy dissipation due to the Coulomb frictional force and the force induced by the path curvature increases much faster than the increase in the friction angle. This effectively controls the gravitational potential energy, and ultimately the particle path, leading to a flatter curve, contrary to the lower friction. 
This is the game between the potential energy and the frictional energy dissipations. With the higher friction, the particle path becomes less and less curved. 
This clearly appears to be the characteristic property that a frictional brachistochrone must possess. However, this is intuitive now because of the involvement of the forces associated with the potential energy and the frictional energy dissipation. Evidently, contrary to the classical brachistochrone, the frictional brachistochrone is asymmetrical. Moreover, unlike in many previous models (Ashby et al., 1997; Lipp, 1997; Hayen, 2005; Covic and Veskovic, 2008; Barsuk and Paladi, 2023), the frictional brachistochrones in Fig. \ref{Fig_1} do not intersect to each other. Furthermore, in contrast to existing models (Ashby et al., 1997; Hayen, 2005), there is no restriction on the frictional parameter $\mu$ in the new model as it dominantly guides the particle path.

\subsection{The straight-brachistochrone}

The important aspect is the friction-architecture of the brachistochrone. As seen in Fig. \ref{Fig_1}, as the friction angle increases, the brachistochrone becomes more and more straight, which is called the straight-brachistochrone. The existence of the straight-brachistochrone is phenomenal. It follows a simple physical principle. As the friction angle increases significantly, because of the reduced net driving force, the particle acceleration decreases resulting in a straight-brachistochrone. 
The crucial point is that, due to the available potential energy, the particle wishes to move down from the incipient position $s_1$, then along some path to the right, to reach the final designated destination $s_2$. However, as the friction begins to play its role, the particle experiences resistance against it moving down. But, as required, it must move to the right to reach the point $s_2$. So, the particle must now orient itself more and more to the right, as it is the only possibility as ruled by the frictional energy dissipation. And, this orientation tendency intensifies with increasing basal friction. In the limit, as the basal friction becomes sufficiently high, the particle tends to move along the straight path joining $s_1$ to $s_2$, if it moves at all. 
However, as the particle potential energy cannot be negative, it cannot move higher than the limiting straight line.
\\[3mm]
Viewed from the mechanical responses between the available potential energy and the frictional energy dissipations, these brachistochrones can be conceived. 
Such fascinating mechanisms are revealed here with the novel equation describing the frictional brachistochrone for a frictional particle traveling along the curved path including its curvature.  

\section{Some applications of the frictional brachistochrone}

Here, I give two applications of the frictional brachistochrone in the optimal design of: (i) the ski jump, and (ii) the bulk granular transportation in process engineering plant, similarly the toboggan run (sled run).  

\subsection{The ski jump}

\begin{figure}[t!]
\begin{center}
  \includegraphics[width=12cm]{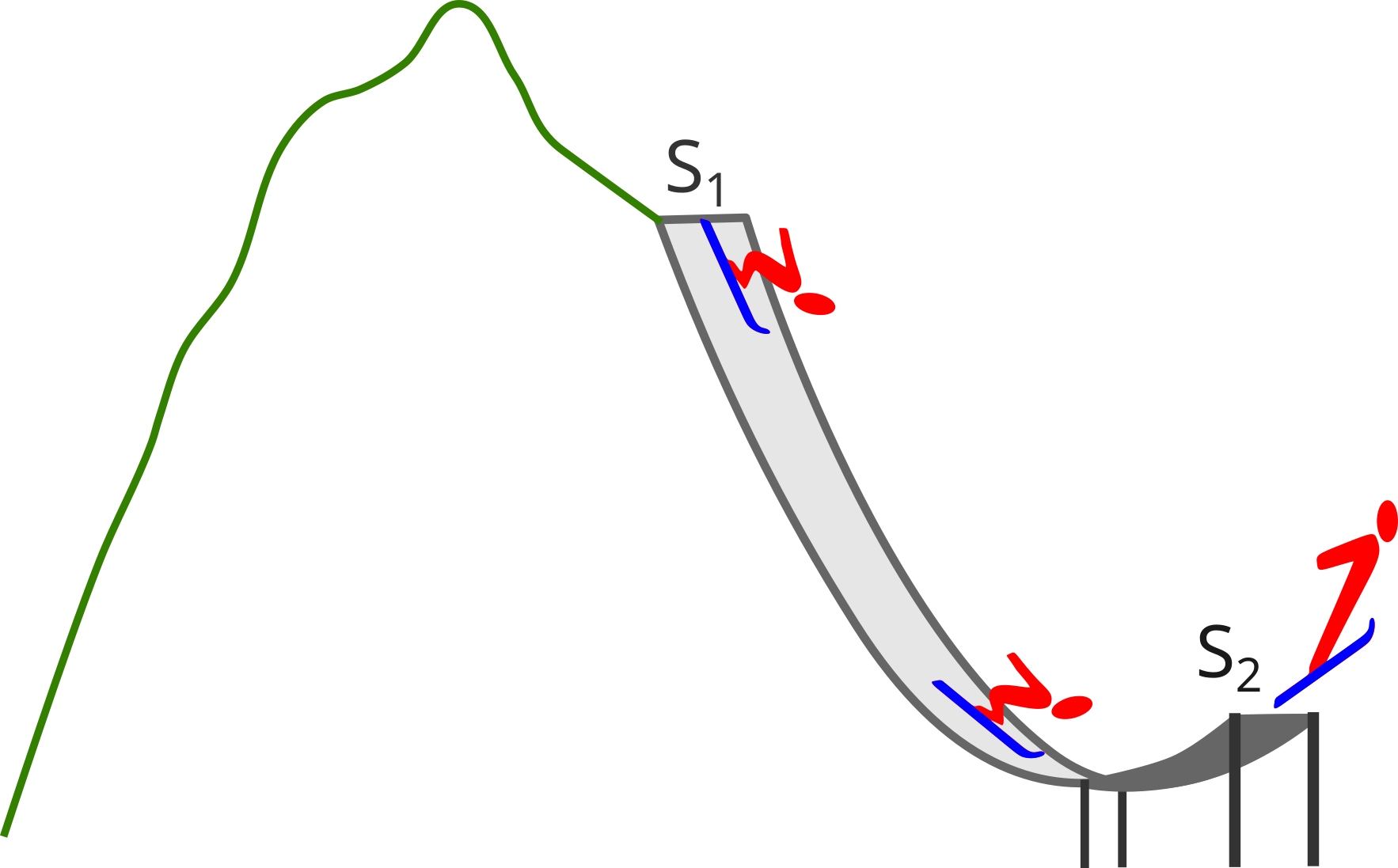}
  \end{center}
  \vspace{-5mm}
  \caption[]{The ski ramp with ski jumper from the initial station $s_1$ at the top until the take-off station $s_2$ at the lower end of the ramp where the ski jumper starts flying.}
   \label{Fig_2}
\end{figure}

The ski jump consists of four parts: run in the ramp, take-off, flight and landing. Here, the main concern is the run in the ramp. The friction between the ski and the snow in the ramp surface plays a vital role in determining the travel time of the skier. The ski jump friction angle is relatively low. Assume it is aimed to design the ramp such that the skier's travel time from the initial position $s_1$ at the top of the ramp to the take-off position $s_2$ at the end of the ramp is minimum (Fig. \ref{Fig_2}). Then, the frictional brachistochrone developed here (Fig. 1) can be used in the construction of such optimal ski jump ramp. The same principle can be applied for the construction of the sand ski jump for which the friction angle is higher than that of the snow ski jump.

\subsection{Bulk granular transportation in process engineering plant}

\begin{figure}[t!]
\begin{center}
  \includegraphics[width=12cm]{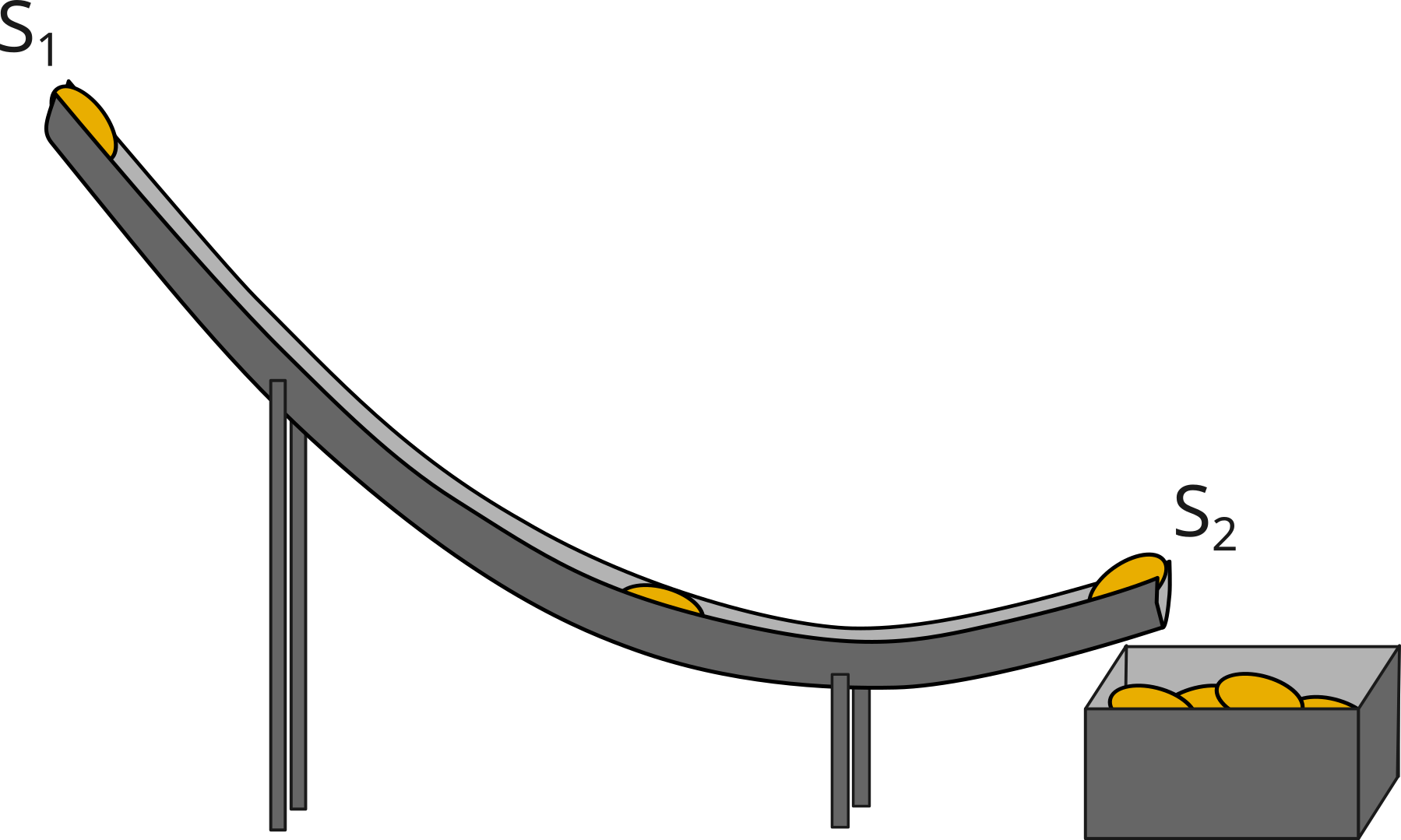}
  \end{center}
  \vspace{-5mm}
  \caption[]{Transportation of the bulk granular material in process engineering plants: the material is set to motion from the initial station $s_1$ at the top of the track (rail) that slides through and reaches the exit station $s_2$ at the lower end where the material is collected.}
   \label{Fig_3}
\end{figure}

Process engineering plants; including the pharmaceuticals, food processing and chemical plants; transport vast amount of granular bulk materials from one position to another during the material in-let, processing and out-let. A scenario is presented in Fig. \ref{Fig_3}, where the material is set to motion from the initial station $s_1$ at the top of the channel (track), that slides through and reaches the exit station $s_2$ at the lower end of the channel where the material is collected. The transportation process primarily depends on bulk material friction between the material and the transportation channel. Production cost is related to the transportation time. The frictional brachistochrone presented here can be utilized to construct the track such that the granular material is transported in the optimal (minimal) time. This, in turn, helps production cost minimization. 
\\[3mm]
As see in Fig. \ref{Fig_1}, the friction determines the optimal time path. A small change in the friction may led to  a substantial alternation of the optimal time path. As the friction between the material and the path is known, the frictional brachistochrone presented here provides the analytical solution to the best design and construction of the ski ramp (Fig. \ref{Fig_2}) and the transportation track (Fig. \ref{Fig_3}) in process engineering.

\section{Summary}

Here, I derived an elegant frictional brachistochrone for a mass point motion of a granular material with the Coulomb frictional energy dissipation together with the curvature of the path. The frictional brachistochrone is represented by a simple ordinary differential equation. Consideration of the system in the path-fitted coordinate made it possible to develop the pleasing representation for the frictional brachistochrone. Structurally and mechanically, this is an astonishing development, genuinely advancing the science of the frictional brachistochrone. An exact, analytical solution to the frictional brachistochrone is constructed. The model can be applied to any frictional particle, without restriction, from exceptionally smooth to incredibly highly frictional. The newly developed frictional brachistochrone clearly demonstrates the significant to dominating effect of the Coulomb friction as it inherently includes the evolving path curvature. It reveals several striking mechanical phenomena. With the higher friction, the particle path becomes less and less curved. This is caused by an interplay between the potential energy and the frictional energy dissipations. In the limit of sufficiently high friction, the existence of the straight-brachistochrone is phenomenal. Some examples of applications of the frictional brachistochrone in the optimal design of the particle path are presented, including the ski jump, toboggan run and the transportation track in chemical process engineering plants.
\\[3mm]
{\bf Acknowledgment:} The financial support is provided by the German Research Foundation (DFG) through the research project: Landslide mobility with erosion: Proof‐of‐concept and application — Part I: Modeling, Simulation and Validation; Project number 522097187. Katharina Boie helped with Fig. \ref{Fig_2} and Fig. \ref{Fig_3}.

\end{document}